%% file: MaxEnt2022_MDPI_templates.tex
\pdfoutput=1
\documentclass[preprints,article,accept,moreauthors,pdftex]{mdpi} %This is especially recommended for submission to arXiv, where line numbers should be removed before posting. For preprints.org, the editorial staff will make this change immediately prior to posting.

\firstpage{1}
\pubvolume{xx}
\issuenum{x}
\articlenumber{x}
\pubyear{2022}
\copyrightyear{2022}
\history{Received: June 2022; Accepted: September 2022; Published: date}
\usepackage{color,xspace}
\usepackage{macros_gpi}

\graphicspath{}

\setitemize{parsep=6pt,itemsep=0pt,leftmargin=*,labelsep=5.5mm}
\setenumerate{parsep=6pt,itemsep=0pt,leftmargin=*,labelsep=5.5mm} 
\setlist[description]{itemsep=0mm}   
\usepackage{environ}
\NewEnviron{myequation}{%
\begin{equation} 
\scalebox{0.95}{$\BODY$}
\end{equation}
}

\Title{Bayesian and Machine Learning Methods in the Big Data era for astronomical imaging} %$^\dagger$}

\Author{Fabrizia Guglielmetti$^{1}$,  Philipp Arras$^{2}$, Michele Delli Veneri$^{3}$, Torsten En\ss{}lin$^{2}$, Giuseppe Longo$^{4}$, Lukasz Tychoniec$^{1}$, Eric Villard$^{1}$} 

\AuthorNames{Fabrizia Guglielmetti, Philipp Arras, Michele Delli Veneri, Tosten Ensslin, Giuseppe Longo, Lukasz Tychoniec, Eric Villard}

\address{%
$^1$ \quad European Southern Observatory, Karl-Schwarzschild-Str. 2, Garching D-85748, Germany
\\
$^2$ \quad Max Planck Institute for Astrophysics, Karl-Schwarzschild-Str.1, Garching D-85748, Germany
\\
$^3$ \quad University of Naples "Federico II" Department of Electrical Engineering and Information Technology, Via Claudio 21, Napoli I-80125, Italy
\\
$^4$ \quad University of Naples "Federico II" Department of Physics "Ettore Pancini", Via Cinthiaaug 21, Napoli I-80126, Italy %; giuseppe.longo@unina.it
}
\corres{Fabrizia Guglielmetti; fgugliel@eso.org}

\firstnote{Submitted to International Workshop on Bayesian Inference and Maximum Entropy Methods in Science and Engineering, IHP, Paris, July 18-22, 2022.} 
\abstract{
  The Atacama Large Millimeter/submillimeter Array with the planned electronic upgrades will deliver an unprecedented amount of deep and high resolution observations.
  Wider fields of view are possible with the consequential cost of image reconstruction. 
  Alternatives to commonly used applications in image processing have to be sought and tested. 
  Advanced image reconstruction methods are critical to meet the data requirements needed for operational purposes. 
  Astrostatistics and astroinformatics techniques are employed.
  Evidence is given that these interdisciplinary fields of study applied to synthesis imaging meet the Big Data challenges and have the potentials to
  enable new scientific discoveries in radio astronomy and astrophysics.
}


\begin{document}
%%%%%%%%%%%%%%%%%%%%%%%%%%%%%%%%%%%%%%%%%%

% For abstract only, no need
% When your abstract is accepted, you can put 
% your main texte here. The same can be used for the final selected paper submission. 

\input{main_texte.tex}

%%%%%%%%%%%%%%%%%%%%%%%%%%%%%%%%%%%%%%%%%%
\reftitle{References}

  \bibliography{biblio/refs}

\end{document}

%% file: main_texte.tex
%%%%%%%%%%%%%%%%%%%%%%%%%%%%%%%%%%%%%%%%%%
\section{Introduction}
\label{sec:Introduction}

The Atacama Large Millimeter/submillimeter Array (ALMA) \cite{wootten} is an aperture synthesis telescope consisting of 66 high-precision antennas.
Sensitive and high-resolution imaging is accomplished employing up to fifty antennas, characterized by 12-meter dishes (12-m Array).
The remaining sixteen antennas compose the ALMA Compact Array (ACA), tailored for wide-field imaging. ACA is characterized by four 12-m antennas
for total power observations and twelve 7-m dishes (7-m Array) for interferometric observations. \\ 
Each antenna is equipped with eight different receiver bands, covering a wavelength range from 3.6 (ALMA band 3) to 0.32 mm (ALMA band 10),
corresponding to a frequency range of 84-950 GHz.\\
Antennas of the 12-m Array can be positioned in a number of different configurations with longest baselines ranging 0.16-16.2 km, which are crucial
in determining the image quality and spatial resolution:
at the highest frequencies in the most extended configurations, the spatial angular resolution reaches 5 mas at 950 GHz \cite{cortes}.
The Array is capable of providing single field and mosaics of pointings.
To make interferometric images, signals from each antenna pair are compared $10^{12}$ times per second within
the ALMA correlator.   
Equipped with a set of correlator modes, ALMA allows both continuum and spectral line observations simultaneously.
\begin{figure}[htb!]
\begin{center}
  \includegraphics[height=4cm,width=7cm]{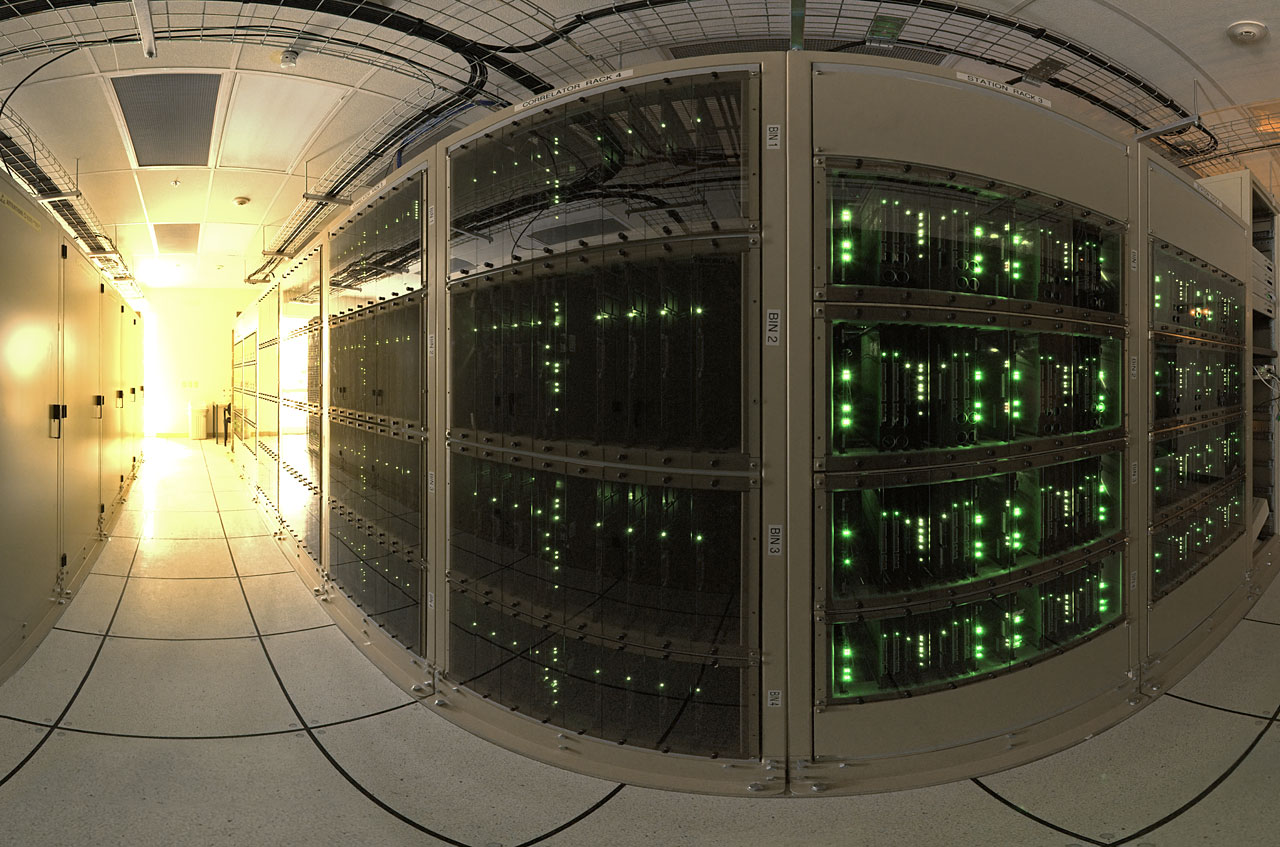} \quad \includegraphics[height=4cm,width=7cm]{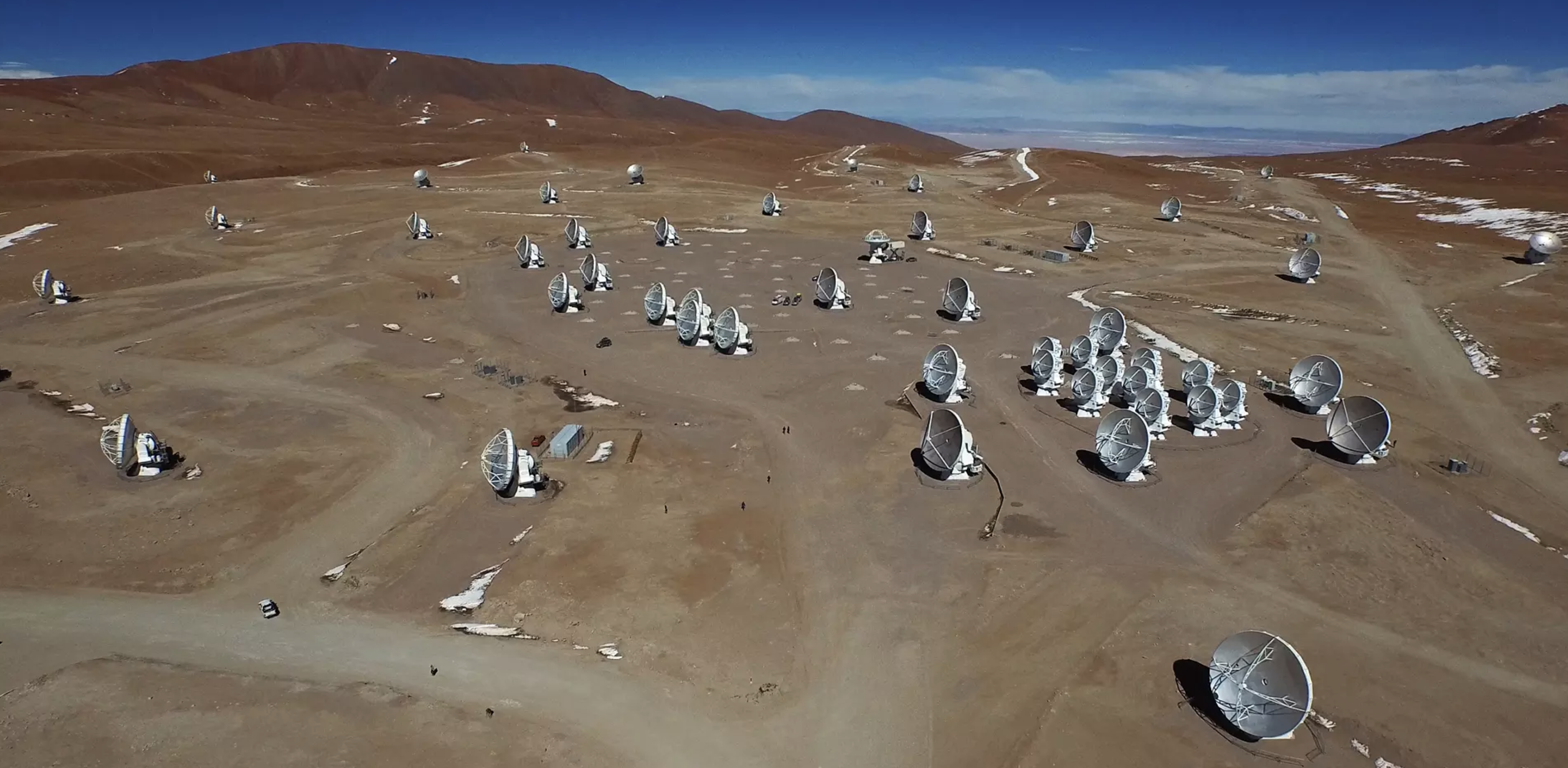}  
\end{center}
\caption{{\em On the left}, the ALMA correlator in the ALMA Array Operation Site (AOS) Technical Building is composed by four identical quadrants with over $134 \times 10^6$  processors, performing up to $17 \times 10^{24}$ operations/s (Image credit: ESO). 
  {\em On the right}, a panoramic view of the ALMA Array, located at an elevation of 5000 m on the Chajnantor Plateau in the Chilean Andes. The AOS is the small building left of picture centre. The tight clump of antennas near the image centre is the ACA (Image credit: JAO).}
  \label{fig01}
\end{figure}

ALMA is undergoing further developments to boost the Full Operation capabilities. 
In the near future, ALMA band 1 \cite{band1} and band 2 \cite{band2} will be installed on each
antenna broadening the receiver bandwidth to cover a total wavelength range of 8.5-0.32 mm (35-950 GHz). 
Moreover, the ALMA2030 Development Roadmap \cite{alma2030} has been approved to keep ALMA as a world leading facility.   
The vision of ALMA2030 accounts for: (1) Broaden the instantaneous bandwidth of the receivers, upgrade the associated
electronics and the correlator to process the entire bandwidth; (2) Improve the ALMA Archive for the end users;
(3) Extension of the maximum baseline length by a factor 2-3.
Another innovative aspect is the design of an array configuration employing all 66 antennas.  
These advancements will enable the following key science drivers: 
Origins of the Planets, Origins of Chemical Complexity (with improved continuum imaging) and Origins of the Galaxies. 
For instance, the study of the Sunyaev--Zel'dovich effect will be enabled to probe the physics of galaxy clusters with the goal of detecting
cluster substructures through high resolution and high-sensitivity observations. 

Currently ALMA is generating 1 TB of scientific data daily. Within the next decade, at least one order of magnitude of increased daily data rate
is foreseen \cite{alma2030}. The planned electronic upgrades (receivers and correlator) will improve ALMA sensitivity and observing efficiency.
In terms of imaging products, ALMA will produce single field and mosaic cubes of at least two orders of magnitude larger than the current 
cube size in the GB regime.
Since the number of observed spectral lines at once will be duplicated, advanced algorithms are needed to provide shorter processing time while
handling larger data volume. Additionally, the imaging algorithms must provide robust and reliable results to reduce human intervention.
Sparse sampling, sky and instrumental responses, pervasive presence of noise increase complexities to the demanding task of image reconstruction.\\
The ALMA development study ``Bayesian Adaptive Interferometric Image Reconstruction Methods'' is providing an initial exploration of concepts that may be of
interest to ALMA development in the long term.   
Using real and simulated data sets, we investigate how to employ Bayesian and Machine Learning techniques to tackle the mentioned challenges.
Specifically for real ALMA data we make use of Science Verification (SV) data. SV is a process by which data quality is assured for scientific analysis.
Observations of a small number of selected astronomical objects are taken with a low number ($\ge 7$) of antennas. 
For ALMA simulated data we make use of the Common Astronomy Software Applications (CASA) \cite{2007casa}, the software package ordinarly used to calibrate, image and
simulate ALMA data.
The performance of a Bayesian and a supervised Machine Learning (ML) techniques is discussed in view of the 
pipeline developments in the ALMA2030 era.
%
%---------------------------------------------------------------

\section{ALMA and the ill-posed inverse problem}
Data contained in ALMA images are affected by the pervasive presence of noise, sparse sampling and instrumental responses. 
The inverse problem of extracting astrophysically interesting information from the observed sky brightness is ill–posed, in the sense that the solution is not unique
or it is not stable under perturbations on the data. Perturbation caused by noise can create large deviations in the solution being sought. 
ALMA interferometric image reconstruction is a demanding task.

The observed data are visibilities of the sky brightness distribution collected by each antenna and correlated for given baseline.
There is one complex visibility (amplitude, phase) for each spectral channel, each correlation and every polarization product.
The visibilities are recorded in integrations with timestamps, one for each baseline. A set of consecutive integrations observing the same celestial direction
(field) forms a scan. A set of scans characterizes an observation. The metadata (or measurement set) contains additional information as known sources
in the field of view, spectral lines, weather (as water vapor in the atmosphere, temperature).  
The data analysis requires the visibilities and metadata concerning the antennas (as positions, diameters), the feeds on the antennas (as sensitivity,
position), the spectral window setup (frequencies, noise, etc). \\
Ideally, i.e.~if the spatial Fourier domain is complete and regularly filled, the inverse Fourier transform of the ensemble of calibrated visibilities provides the transformation function to move from the spatial frequency domain (in units of
flux density $[Jy]$) to the image plane (in units of surface brightness $[Jy/beam]$) \cite{SIRA2}, \cite{MaxEnt2019}: $I^{D}=I_{db} \ast {I \cdot A}$.
In practice, the spatial Fourier domain is partially and irregularly filled. The inverse problem becomes Fourier Synthesis inverse problem. 
The resulting dirty image $I^{D}$ is corrupted by the incomplete sampling and the instrumental point spread function (dirty beam) $I_{db}$, with the dirty beam being a function of the uv sampling.
The dirty beam is the instrumental response of an observation and it is characterized by strong sidelobes corrupting the image.
Deconvolution of the dirty image $I^{D}$ from the dirty beam $I_{db}$ results in an image with a flux distribution corrupted by an additional instrumental response, named primary beam $A$. 
The primary beam expresses the sensitivity of the instrument as a function of direction, with typically being most sensitive at the phase centre and dropping off away from the pointing
direction. The primary beam effects are, traditionally, removed by dividing the deconvolved image by an average primary beam pattern.
The effect of this process in the final corrected image is increased image noise towards the map edge. 
Sampling and point spread functions vary with the observation setup.

\subsection{RESOLVE for Bayesian signal inference}
Given the ill-posedness of the imaging task, the {\rm RESOLVE} algorithm \citep{2016resolve,2016fastresolve,2018resolve} reconstructs images from the detected signal assuming a Gaussian
likelihood and a Fourier space response function. The detected signal $d$ is described by the measurement equation $d=Re^{s} +n$, with $s$ describing the real sky signal corrupted by the instrumental
response $R$ and additional noise $n$ contaminating the real sky signal.
The statistical description of the real brightness distribution occurs by inferring the most probable signal $s$ given the measured visibility function $d$.
Information field theory \citep{IFT, IFT2} is used connecting statistical field theory and Bayesian inference. The posterior probability density function of the celestial signal given the observed data $P(s|d)$ is related to 
the information Hamiltonian $H(d,s)$:
\begin{equation}
P(s|d)=\frac{e^{-H(d,s)}}{{\mathcal Z}(d)},
\end{equation}
where the partition function ${\mathcal Z}(d) := \int {\mathcal Ds P(s,d)}$ and ${\mathcal H(s,d)}:=-log{\mathcal P(s,d)}$. $\mathcal \int Ds$ is the path integral defined as the continuum limit of the
product of integrals over every image pixel $\int \prod_{i} ds_{i}$ \cite{2018resolve}. Additionally, the technique is capable of inferring the signal covariances of the sky brightness distribution, 
the noise level of each data point and the power spectra of $s$. Products of the data analysis are the reconstructed signal and uncertainty maps, the power spectrum and estimates of the initial input parameters. \\
Originally designed to detect spatially extended source brightness distributions \cite{2016resolve}, {\rm RESOLVE} was subsequently delivered with a speed up procedure while introducing
a Bayesian estimation of the measurement uncertainty of the visibilities into the imaging \cite{2016fastresolve}. In \cite{2018resolve}, the optimization procedure of the technique was refined
allowing the noise level of each data point to be learned simultaneously with the map reconstruction. The main advantage is convergence speed up to the optimal solution in a high-dimensional space.
Calibration and imaging procedures were introduced as one unique algorithm by \cite{ediss27827}, allowing for error propagation. 
{\rm RESOLVE} comes with its own python library (see e.g.~\cite{nifty}) and the software is  continuously evolving.
{\rm RESOLVE} proved to be superior for extended and faint emission detection on the Very Large Array radio data with respect to the traditional {\rm CLEAN} method \cite{Arras_2021}. {\rm CLEAN} is a {\rm CASA} task
composed by several operating modes, allowing for the generation of images from visibilities and the reconstruction of a sky model \citep{2007casa, MaxEnt2019}. Applications
of {\rm RESOLVE} on ALMA data and simulated ALMA data can be found in \cite{MaxEntlukasz}, this conference proceedings, where a comparison with {\rm CLEAN} is provided.  

\begin{figure}[htb!]
 \centering
 \begin{tabular}{cc}
        \vspace{-5cm} \includegraphics[width=7cm]{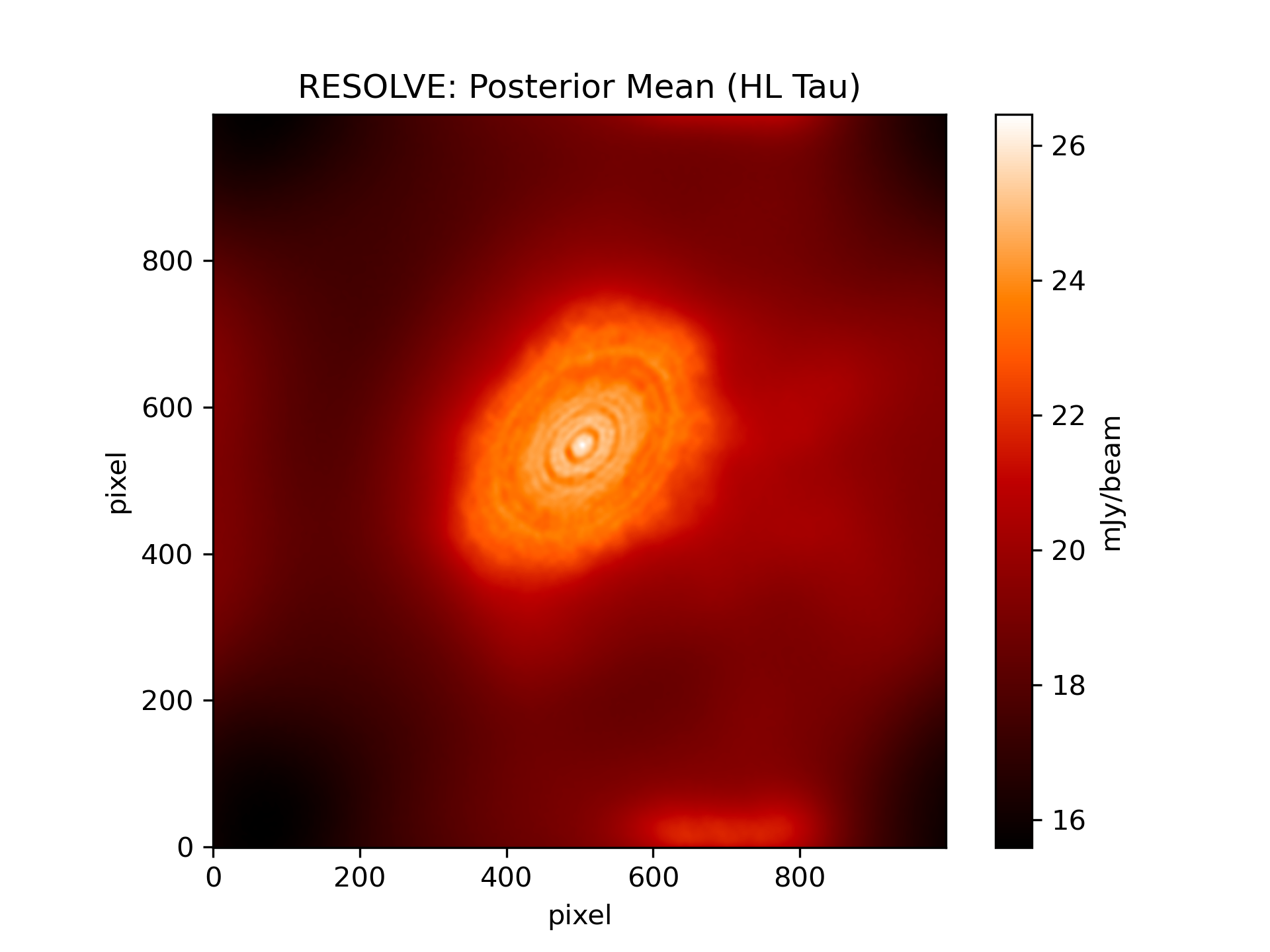}   \\ 
       \includegraphics[width=7cm]{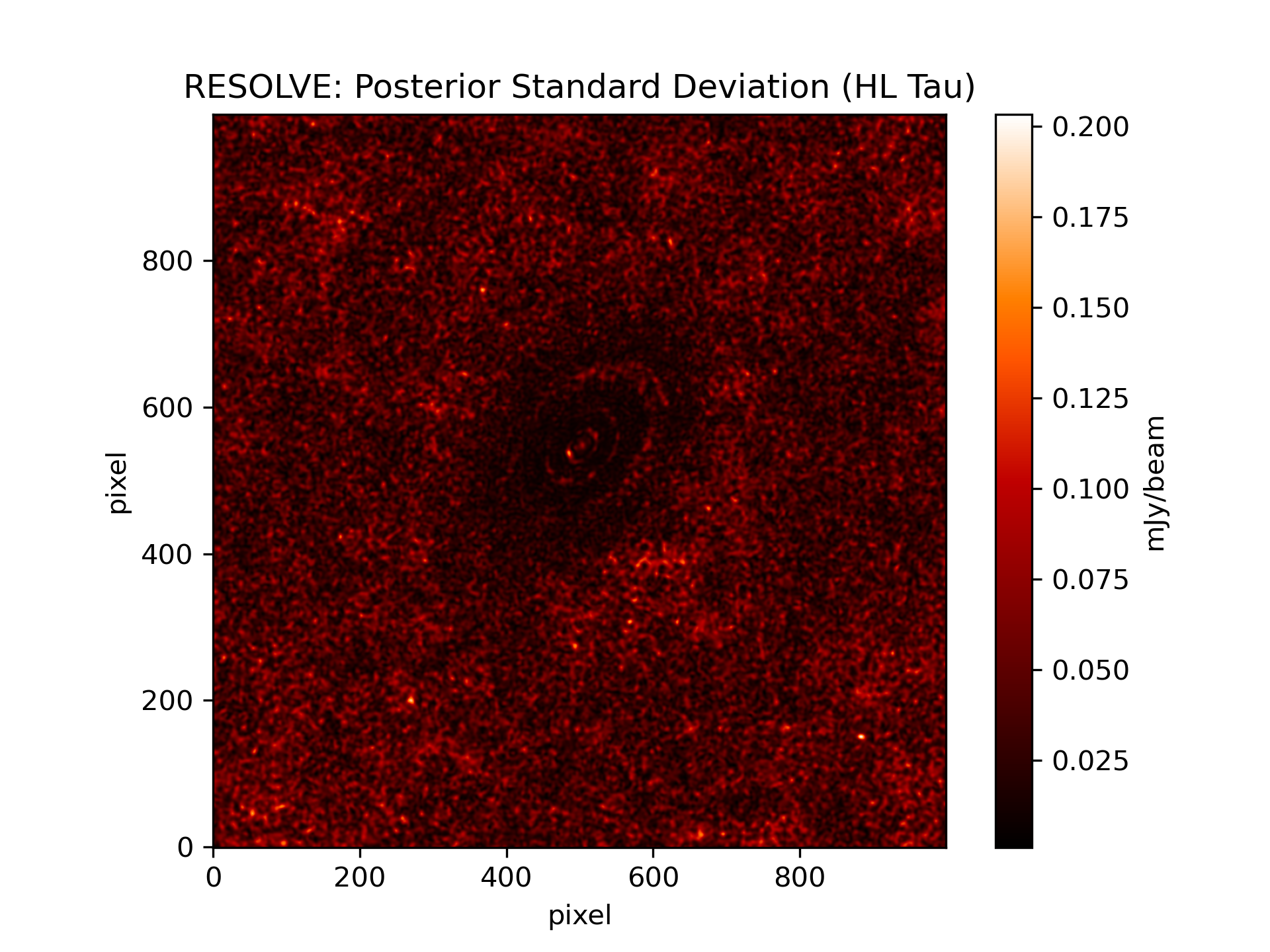} &
       \includegraphics[width=5cm]{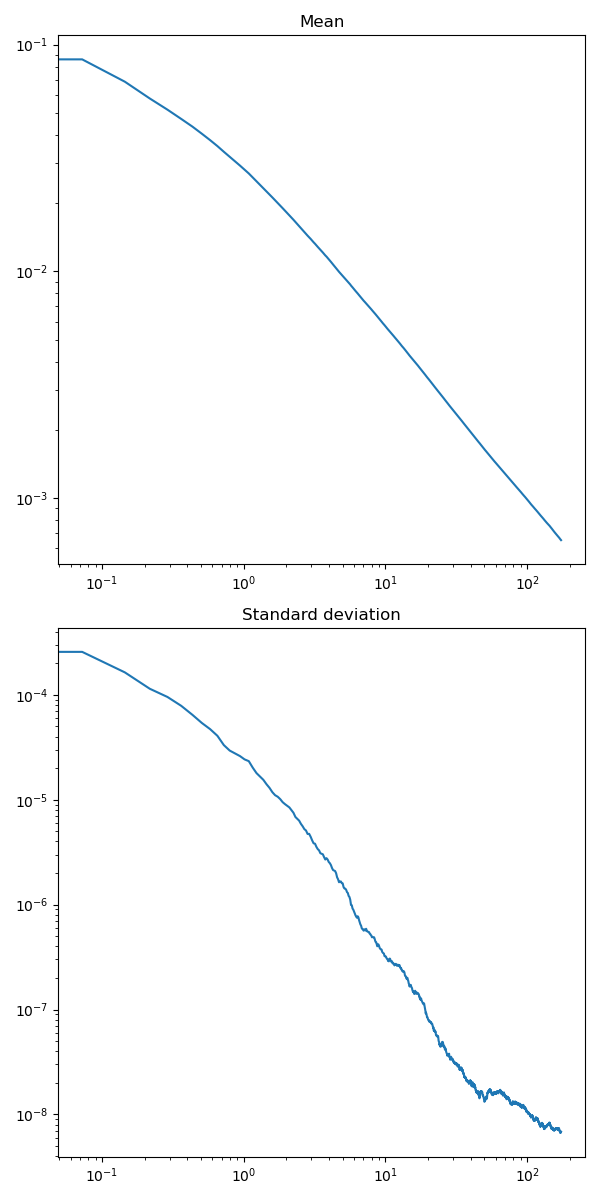} 
 \end{tabular}
  \caption{Application of {\rm RESOLVE} on SV data, HL Tau, continuum image at 1.3 mm (233 GHz), using only one out of four 1.8275 GHz
    spectral window composed by 128 channels.
    {\em On the left}, image reconstruction of HL Tau in units of [mJy/beam] indicating the posterior mean (upper) and the relative pixel-wise posterior uncertainty 
    (lower). {\em On the right}, the estimated mean spatial correlation, or posterior power spectrum of the reconstructed image (upper) and its uncertainty (lower).}
\label{fig04}
\end{figure}

In Fig.~\ref{fig04}, an application of {\rm RESOLVE} on ALMA SV data is shown. HL Tau ($04h 31m 38.4s +18^{\circ} 13^{'} 57^{''}$, J2000) \cite{2015HL}
was observed in 2014 with long baselines (up to 15.2 km) employing about 30 antennas. This science target was observed for 4.5 hours allowing for
good sampling and reaching a very high angular resolution ($0.035 \times 0.022$ arcsec). Although {\rm RESOLVE} is employed on a quarter of the available
band 6 data, 
this application is showing the remarkable morphology of the protoplanetary disk, with patterns of alternating bright and dark rings,
confirming the finding in \cite{2015HL}. {\rm RESOLVE} is a Bayesian algorithm providing not a single reconstructed image, but a probability distribution
of all possible image configurations. It follows that the image reconstruction, shown in the upper left, is the posterior mean representation. The uncertainty map of the
reconstructed sky image by {\rm RESOLVE} is provided  (lower left image).  {\rm RESOLVE} learns from the data the power spectrum with the sky brightness. 
The posterior mean power spectrum of the log-sky brightness distribution is shown in the upper right image: the power spectrum $P_{S}(k)$ of the process that generated the signal $s$ as a function of spatial frequency $k$ \cite{2016resolve}. The estimated uncertainty on the mean power spectrum is shown below (lower right image).\\
Because of the algorithm design, {\rm RESOLVE} provides the tools for data combination (data fusion). This technique has the potential to achieve successfully the data combination challenge when the signal of all 66 ALMA antennas will form one unique data set as in view of the ALMA2030 roadmap. 

\subsection{Deep learning for fast image reconstruction}
\begin{figure}[ht] 
\begin{center}
\includegraphics[height=7cm,width=12cm]{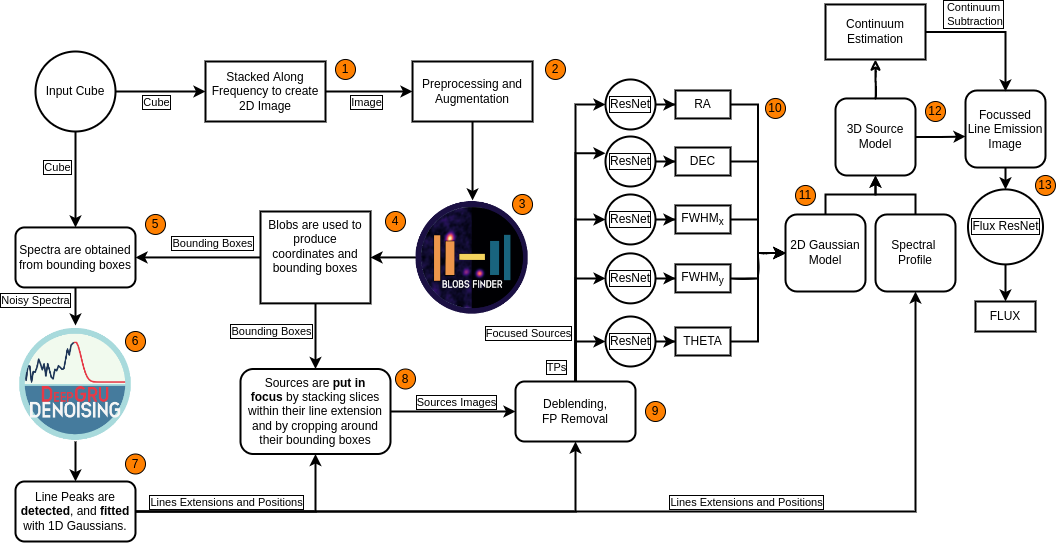}
\end{center}
\caption{Schematic view of the DL pipeline used within the 3D detection and characterization of ALMA sources \cite{delliveneri1}, with the numbers showing the logic flow of the data.}
\label{fig02}
\end{figure}

Supervised machine learning (ML) techniques provide powerful tools to quickly analyse large datasets and to support for easier collection and storage.
In particular, deep learning (DL) algorithms have been widely applied in many areas of astronomy due to the capacity to automate feature selection \cite{2019Loredo}.\\
The DL Pipeline \cite{delliveneri1} ({\it paper submitted}) is developed to detect and characterize sources in ALMA cubes.
The idea is to learn the underlying features from the ALMA dirty image cubes ($I^{D}$). 
An architecture of several ML methods compose the DL Pipeline (Fig.~\ref{fig02}): Blobs Finder, Deep GRU and ResNet. 
Convolutional architectures (Blobs Finder and ResNets) are used to process spatial information. Recurrent Neural Network (Deep GRU) is employed to process
sequential information. 
\begin{figure}[ht] 
\begin{center}
\includegraphics[height=6cm,width=6cm]{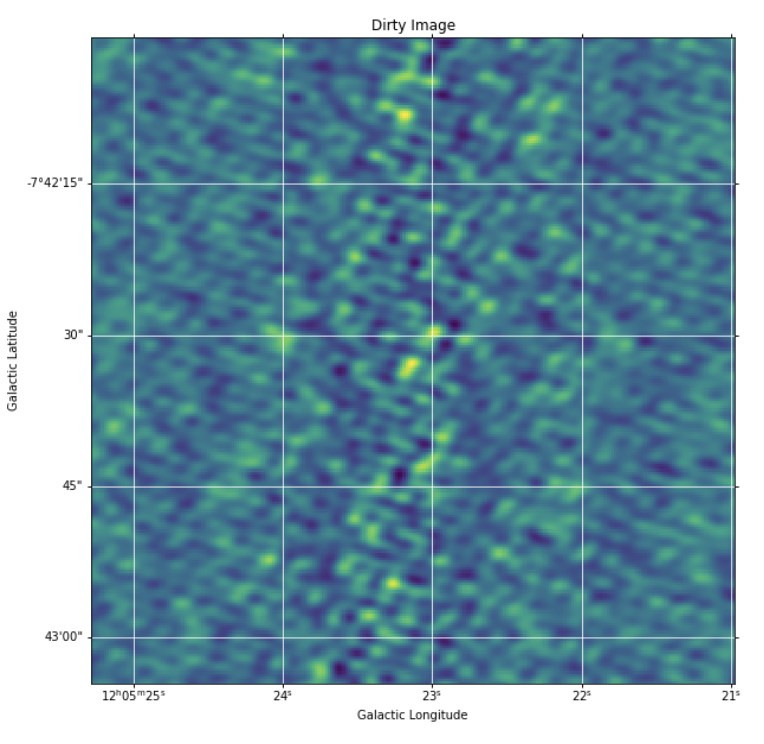} \quad \includegraphics[height=6cm,width=6cm]{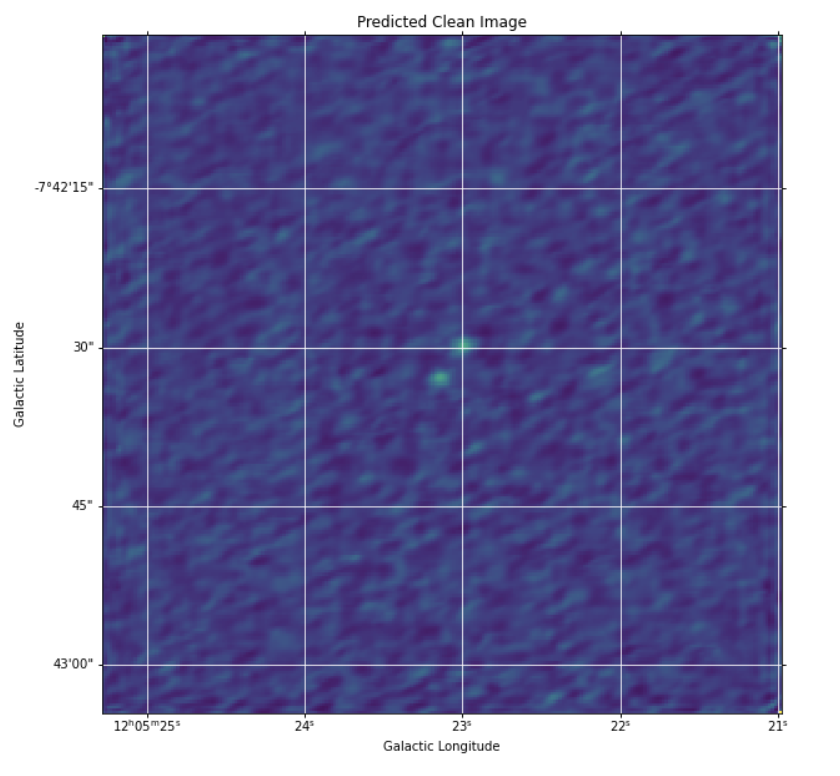}
\end{center}
\caption{{\em On the left}, [CII] line emission observed at 0.9 mm (334 GHz): integrated intensity map from ALMA dirty cube in BR1202-0725, a binary system observed edge-on and with the galaxies moving
  along the line of sight. {\em On the right}, the predicted image with the DL Pipeline.}
\label{fig03}
\end{figure}
\begin{figure}[ht] 
\begin{center}
  \includegraphics[height=5.3cm]{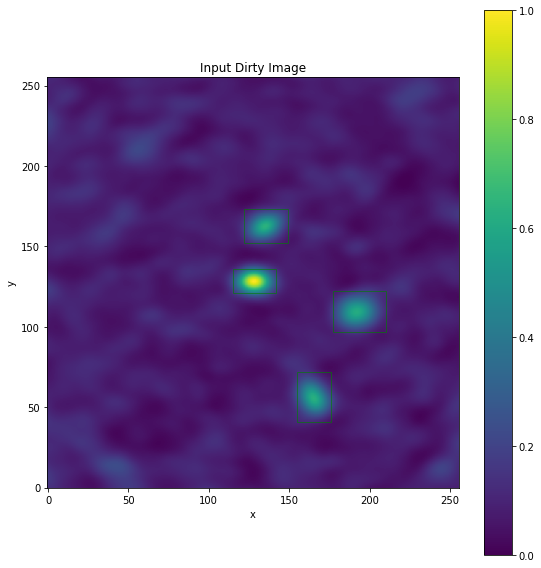} \includegraphics[height=5.3cm]{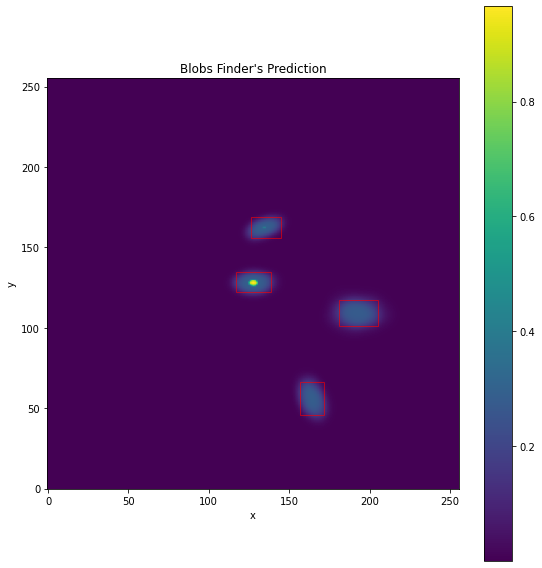} \includegraphics[height=5.3cm]{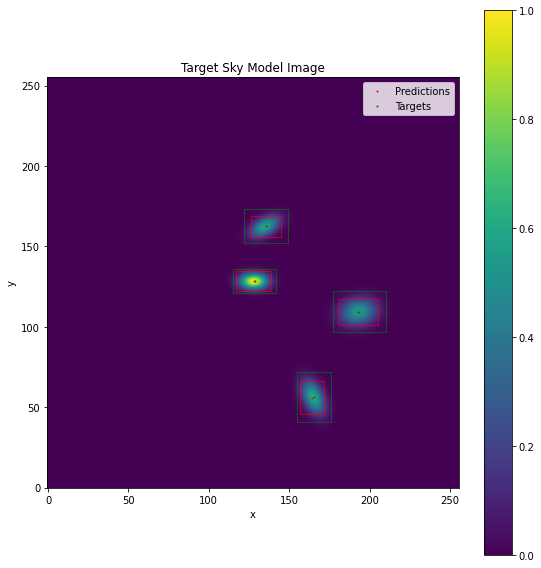} 
\end{center}
\caption{Application of Blobs Finder (part of the DL Pipeline) to ALMA simulated dirty images. {\em On the left, center and right}, the simulated ALMA dirty cube,
  Blobs Finder prediction and Sky model, respectively. These are integrated intensity maps. The red boxes indicate the lines found by Blobs Finder. Spectral analysis is performed on each detected line by ResNets in a subsequent stage.
}
\label{fig05}
\end{figure}

Blobs Finder is a 2D Deep Convolutional Autoencoder. Blobs Finder is used to detect sources in the ALMA images. 
The Autoencoder is composed by an input encoder, performing image compression into a lower dimensional latent space, and an output decoder that decompresses from the latent space.
The latent space simplifies the data representation for the purpose to find patterns. 
Image reconstruction occurs by decompressing the lower dimensional latent space, that forces the model to learn the most relevant features in the images: sources, point spread function, noise.  
Blobs Finder addresses the deconvolution problem: $ I^D(x,y) = R \ast I(x,y) + \epsilon$, where $I^D(x,y)$ is the dirty image produced integrating the ALMA
dirty cube along the frequency, $I(x,y)$ represents the true sky representation,
$R$ is the intrumental response and $\epsilon$ is any additional noise.\\
Each detected source is analysed in frequency space in the dirty cube with the Deep Gated Recurrent Unit (GRU). Deep GRU is a Recurrent Neural Network, suited for
sequential data analysis. Deep GRU allows for internal memory about the received input with a consequent high predictive power.
It denoises the spectra extracted from the detected sources in search for (emission or absorption) lines.
Spurious signal is removed and each detected source goes through the final step. \\
The Residual Neural Networks (ResNet), a class of Deep Convolutional Neural Networks, are used to predict source parameters, including flux estimates.

ALMA SV interferometric data are used to test the capabilities of the DL pipeline. In Fig.~\ref{fig03}, left, the field targeting BR1202-0725 \cite{2013carniani} observed with 18 12-m diameter antennas in 2012 is shown.
The target was observed for a total time of 25 minutes during stable weather conditions and employing a maximum baseline of 280 m.    
In Fig.~\ref{fig03}, right, the detected BR 1202-0725 system with the DL pipeline is shown. The submm galaxy (north) and the quasar (south) are visible ($z\sim 4.7$) and separated by the noise and point spread function effects. The physical properties of the submm galaxy and of the quasar are summarized in table 1 of \cite{2013carniani}. The source fluxes derived by the DL pipeline agree
with the ones reported in \cite{2013carniani}. The computing time for the image restoration occurred in $\sim 35 \mu s$.

ALMA cubes have been simulated to train the algorithm in detecting galaxies.
At least 1000 cubes are generated of which 4/5 are used to train the models, 1/10 to validate the models and 1/10 to test the models
performance. In Fig.~\ref{fig05}, an example of ALMA simulated dirty cube with four galaxies is shown (on the left).
In this example, emission lines are randomly injected in the synthetic ALMA cubes.
The image reconstruction obtained with Blobs Finder is at the center, providing a solution very close to the real simulated sky (on the right).   

The algorithm is going to be trained in detecting other kinds of celestial signals and to be tested in more complex environments, e.g.~serendipitous detection
of obscured quasi stellar objects. 
Advancements on uncertainty quantification are foreseen, modifying the networks and the loss functions to allow for measurement error propagation through the pipeline.

\section{Conclusions}
Through this study we provide the initial exploration of concepts in the search for novel imaging techniques applicable on large data volume, thus enabling  
efficiency improvements in data processing while requiring the least amount of human intervention. 
We employ two distinct software to analyse ALMA data in view of the challenges arised by the ALMA2030 development roadmap. The two techniques are different
in nature, one based on astrostatistics the other on astroinformatics \cite{2019aneta}. Both techniques demonstrated to be equipped by essential strengths
and by required features the Big Data era is longing. 

{\rm RESOLVE} is a robust algorithm, founded on a principled method. It is designed for the detection of diffuse emission. Complex structures in the celestial signal
and point-like sources are well detected. The input parameter values are initialized and estimated by the data during the optimization to the most probable image
configuration. The reconstructed images provide for a reliable solution with no need of extra human intervention. {\rm RESOLVE} was applied on ALMA continuum images.
Applications of the technique on ALMA aggregate continuum and cubes are planned. Although {\rm RESOLVE} is computing expensive, the algorithm delivers in addition
to the reconstructed image other informative products (as uncertainty map, power spectrum, final values of the estimated input parameters). These products have
the potentials to lay the foundations for designing a fully automated pipeline. 

The DL pipeline demonstrated high image fidelity and high-performance computing for image reconstruction on ALMA data cubes.
The technique is applied on ALMA dirty cubes, learning from the image the celestial sources, the noise, the instrumental point spread function. 
It allows for extreme data compression by leveraging both spatial and frequency information.
In the near future, the DL pipeline will be applied on continuum images and trained and tested on a large variety of celestial signals.  
Nonetheless, astroinformatics has the potential to revolutionise data management in Science Archives.  
ALMA is currently producing roughly 500TB worth of raw-data and reduced data products per year.
Currently, 5\% of the total data volume in the archive is occupied by images.
The DL Pipeline may allow to create images on user demand with a one-click system through a web-interface.
Moreover, catalogues creation of stored data per ALMA Cycle is feasible in an automated fashion. 
  
As a conclusive remark, based on the current investigations, {\rm RESOLVE} is the algorithm of choice for robust diffuse emission and faint source detection while the DL Pipeline implemented within {\rm CLEAN} will meet the feasibility requirements of ALMA2030 performance algorithm.      
Because of the planned upgrades, ALMA image analysis strives for algorithms capable of discovering through the data, to adapt when given new data and to get the most out of the data.      

\funding{ This research is supported by an ESO internal ALMA development study investigating interferometric image reconstruction methods.}